\documentstyle[amsfonts,prl,aps]{revtex}

\input epsf

\renewcommand{\vec}[1]{\mbox{\boldmath$#1$}}

\begin{document}

\twocolumn

\title{Spin Excitations in a Fermi Gas of Atoms}

\author{B. DeMarco \cite{adr1} and D. S. Jin \cite{adr2}}

\address{JILA, \\ National Institute of Standards and Technology
and University of Colorado, Boulder, CO 80309}

\date{\today}

\maketitle

\begin{abstract}
We have experimentally investigated a spin excitation in a quantum
degenerate Fermi gas of atoms.  In the hydrodynamic regime the
damping time of the collective excitation is used to probe the
quantum behavior of the gas.  At temperatures below the Fermi
temperature we measure up to a factor of 2 reduction in the
excitation damping time.  In addition we observe a strong
excitation energy dependence for this quantum statistical effect.
\end{abstract}


\narrowtext

Studies of elementary excitations have proven to be an effective
tool for investigating the dynamics of dilute, ultracold atomic
gases\cite{review}. These systems are fertile testing grounds for
many-body quantum theory because the interactions, which arise via
collisions between the constituent atoms, can be described from
first principles. For fermions there are essentially no
interactions in a single component gas at ultralow temperature.
However an interacting Fermi gas was recently realized using
magnetically trapped $^{40}$K atoms in two different spin
states\cite{demarco2001}. The collective excitation behavior for
such a mixed spin component Fermi gas is rich and has been
investigated theoretically in both the collisionless and
hydrodynamic regimes
\cite{Vichi1999,Bruun1999,Amoruso2000,Vichi2000,Bruun2000a,Minguzzi2001}.
Studies of spin excitations may also serve to expose the onset of
a superfluid phase in the Fermi gas
\cite{Baranov2000,Bruun2000b,Minguzzi2001}. In the work presented
here, we use a dipole (``slosh") spin excitation in the
hydrodynamic regime to probe a Fermi gas of magnetically trapped
$^{40}$K atoms.  This type of excitation, which was previously
shown to be sensitive to interactions \cite{gensemer2001}, is
shown here to probe quantum statistical effects in dynamics of the
Fermi gas.

The production of an interacting Fermi gas of atoms follows our
previous experimental procedure \cite{demarco2001,demarco1999}.
Fermionic $^{40}$K atoms are magnetically trapped in two spin
states (the $m_f=9/2$ and $m_f=7/2$ Zeeman states in the $f=9/2$
hyperfine ground state) in order to permit the s-wave collisions
necessary for evaporative cooling to ultralow temperature.  As in
all experiments with atomic Fermi gases
\cite{demarco1999,hulet2001,paris2001}, two components are
required because collisions between identical fermions are
forbidden by symmetry at the temperatures of interest. The
interactions therefore arise in the gas from collisions between,
but not within, the $m_f=9/2$ and $m_f=7/2$ components.

In the magnetic trap, the $m_f=9/2$ and $m_f=7/2$ components have
slightly different single-particle harmonic oscillator frequencies
because of a small disparity in their magnetic moments.  This
small difference in the bare harmonic oscillator frequencies
determines the collision rate necessary to reach the hydrodynamic
regime for the type of excitation used in this work
\cite{gensemer2001}. For the measurements described in this
Letter, the single-particle oscillator frequencies for the
$m_f=9/2$ component were 256 Hz radially and 19.8 Hz axially
\cite{note}. The bare trap frequencies for the $m_f=7/2$ component
are reduced from these values by a factor of $\sqrt{7/9}$.

The excitation studies used a gas of 0.7 to 3.3 million atoms, in
a nearly equal mixture of the two-spin states, cooled to a
temperature between 0.24 and 2.0 $\mu$K. A dipole, or slosh,
oscillation was excited by shifting the trap center for
approximately half of an axial period (28~ms) along the axial trap
direction as described in Ref. \cite{gensemer2001}. The resulting
motion of the gas, involving center-of-mass oscillation of each
component about the trap center, was allowed to evolve in the
magnetic trap.  Before imaging the gas, the magnetic trap was
quickly turned off so that the gas expanded ballistically for 16
ms. During the expansion, the two components were spatially
separated by a magnetic field gradient
\cite{demarco2001,ketterle1998}.  The number, temperature, and
center of mass of both components were determined from independent
fits to each absorption image taken after expansion. The dynamics
of the dipole excitation were mapped out by varying the evolution
time in the magnetic trap before release.

The center-of-mass positions of both components were recorded as a
function of time and then fit simultaneously to exponentially
damped harmonic motion in order to extract the normal mode
frequency and damping rate.  The axial separation between the
centers of mass of the two spin components varied during the
dipole oscillation.  Therefore, the axial magnetization in the
center-of-mass frame of the cloud oscillated in space and time.
This type of excitation is reminiscent of the well known spin
waves in quantum fluid and solid state systems, and is similar to
the spin waves observed in spin-polarized hydrogen gases
\cite{spinwaves}. The existence of a spin excitation requires a
spin dependent interaction.  In our case, spin dependent forces
arise from the magnetic trapping potential and also from the
quantum nature of the collisions.  Even at temperatures well above
the degenerate regime for bosons or fermions, the quantum
statistics distinguishes collisions between atoms in the same
internal state and those between atoms in different internal
states.  One might expect that collisions between atoms in
different spin states would strongly damp a spin excitation, which
must involve relative motion of the two spin components. However,
the normal mode that we observe in the hydrodynamic regime
consists of nearly in-phase motion of the two spin components such
that the amplitude of the variation in magnetization is relatively
small.

Before probing the quantum Fermi gas, we explored the spin
excitation in the non-degenerate regime \cite{gensemer2001}. While
the frequency of the spin excitation was nearly independent of the
collision rate $\Gamma_{\sf coll}$ in the gas, the damping time
depended linearly on $\Gamma_{\sf coll}$. In Fig. 1 we plot the
measured exponential damping time $\tau$ for the spin excitation
vs. the average collision rate per particle, $\Gamma_{\sf coll}$.
The collision rate per particle is defined by $\Gamma_{\sf
coll}=2n\sigma v/(N_{9/2}+N_{7/2})$, where the density overlap
$n=\int\,n_{9/2}(\vec{r})\,n_{7/2}(\vec{r})\,d^3\vec{r}$, the
total number of atoms $N_{9/2}+N_{7/2}$, and the mean relative
speed $v=4\sqrt{\frac{k_BT}{\pi m}}$ for a collision between a
$m_f=9/2$ and $m_f=7/2$ atom are determined from Gaussian fits to
the absorption images.  The s-wave cross section is given by
$\sigma=4\pi a^2$, where the triplet scattering length for
$^{40}$K is $a=169a_0$ \cite{scatt} ($a_0$ is the Bohr radius).
The data in Fig. 1 are taken in the non-degenerate regime, with
the temperature $T$ of the gas greater than the Fermi temperature
$T_F$ for either spin-state component. The Fermi temperature
depends on the number of atoms $N$ and the radial and axial
harmonic oscillator frequencies $\omega_r$ and $\omega_z$ through
$T_F=\hbar\,(6\,\omega_r^2 \omega_z N)^{\frac{1}{3}}/k_b$
\cite{rokshar}.

\begin{figure}
\begin{center}\epsfxsize=3 truein \epsfbox{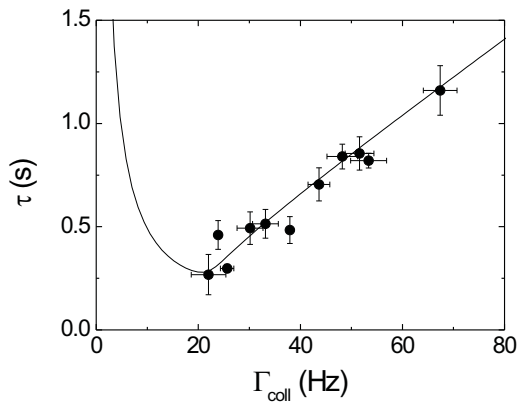}
\end{center}
\caption{Damping time of the spin excitation. The exponential
damping time $\tau$ of the nearly in-phase normal mode is plotted
versus the collision rate per particle, $\Gamma_{\sf coll}$. A fit
of the data to a classical kinetic model is shown by the solid
line.  The data shown here are taken in the non-degenerate regime,
$T/T_F>1$, and with sufficiently high $\Gamma_{\sf coll}$ to be in
the regime of hydrodynamic behavior. The vertical error bars in
$\tau$ represent the uncertainty from the fits to the
center-of-mass motion of the gas, while the horizontal error bars
represent variation in $\Gamma_{\sf coll}$ during the oscillation
measurement. \label{fig1}}
\end{figure}

As expected, the damping time increases linearly with the
collision rate in the hydrodynamic regime \cite{nozieres}. The
increase in damping time, or equivalently reduction in damping
rate, is due to the fact that the higher collision rate more
firmly establishes the collective excitation.  The data in Fig. 1
are fit to a classical kinetic model\cite{gensemer2001}, where the
spin components are treated as two harmonic oscillators coupled by
collisional viscous damping. The data agree well with the model,
which includes a scaling constant on the collision rate axis as
the only fit parameter. The best fit value for this scaling factor
is within our estimated 50\% uncertainty in determining the number
of atoms.

The quantum nature of the gas is revealed through changes in the
excitation dynamics as the gas is cooled below $T_F$.  The damping
time of the dipole oscillation is measured for an equal mixture of
$m_f=9/2$ and $m_f=7/2$ atoms as the temperature of the gas is
varied through forced evaporative cooling.  The emergence of
quantum behavior below $T_F$ is observed by comparing the measured
damping time $\tau$ to the classical prediction $\tau_{\sf class}$
in the hydrodynamic regime. The measured spin excitation damping
time is shown in Fig. 2. The classical prediction is determined by
the value of $\Gamma_{\sf coll}$ inferred from the measured $n$,
$v$ and $N_{9/2}+N_{7/2}$ and the fit shown in Fig. 1.

\begin{figure}
\begin{center}\epsfxsize=3 truein \epsfbox{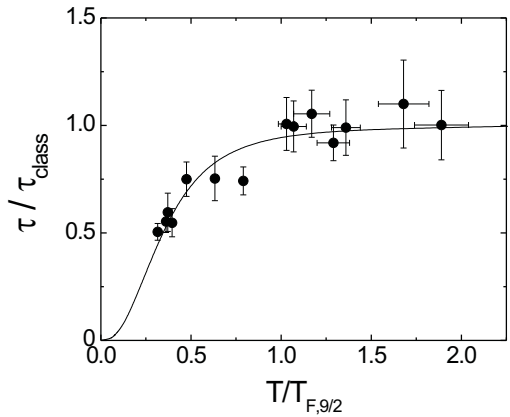}
\end{center}
\caption{Effect of quantum degeneracy on the spin excitation
damping time. At low $T/T_F$ the measured damping time $\tau$ is
reduced compared to the classical expectation $\tau_{\sf class}$.
The classical expectation $\tau_{\sf class}$ is obtained from the
measured collision rate per particle using the fit shown in Fig.
1. The y-axis error bars in this figure and in Fig. 3 include the
uncertainties from the fit to the center-of-mass motion, the
variation in $\Gamma_{\sf coll}$ during the oscillation
measurement, and the fit from Fig. 1. Variation in the temperature
during the oscillation measurement is represented by the error
bars for $T/T_{F,9/2}$.  The data agree well with the theoretical
prediction from a quantum kinetic calculation (solid line) of the
effect of Pauli blocking on the collision rate. \label{fig2}}
\end{figure}

At low $T/T_F$ we observe that the damping time decreases
significantly compared to the classical expectation. This change
in the damping arises from a quantum statistical suppression of
collisions.  In the degenerate Fermi gas low energy quantum states
have an occupancy approaching one.   This fact combined with the
Pauli exclusion principle suppresses the elastic collision rate
through a reduction of allowed final states.  As seen in Fig. 1,
lowering $\Gamma_{\sf coll}$ in the hydrodynamic regime results in
a shorter damping time (or equivalently a higher damping rate).
Thus by reducing $\Gamma_{\sf coll}$, Pauli blocking hinders the
ability of the collective excitation to propagate in the quantum
regime.  Ultimately, in the zero temperature limit one expects the
gas to reach the collisionless and zero-sound regime.

The data in Fig. 2 are compared with the prediction of a quantum
kinetic calculation of the $\Gamma_{\sf coll}$ suppression due to
Pauli blocking \cite{holland2000}, shown by the solid line. The
calculation includes the difference in trap frequencies for the
$m_f=9/2$ and $m_f=7/2$ components and assumes an equal mixture of
atoms in the two spin states.  For comparison with the
measurements, we have plotted the calculated quantum collision
rate normalized by the collision rate calculated without Pauli
blocking.  The nearly factor of 2 reduction in $\tau$ observed at
our lowest $T/T_F$ agrees with the theory.

Pauli blocking of collisions was seen previously in measurements
of rethermalization rates in the two-component Fermi gas
\cite{demarco2001}.  Although both the data in Fig. 2 and the
previous results can be understood in terms of the very general
quantum effect of Pauli blocking, it is important to note that the
two experiments probe very different dynamics.  In fact, the sign
of the measured effect in the two experiments is opposite. The
measured rethermalization time increased in the quantum regime,
while the measured damping time shown in Fig. 2 decreases in the
quantum regime.

The damping rate for the collective excitation also exhibits a
strong amplitude dependence in the quantum regime. The dependence
of the exponential damping time on the center-of-mass oscillation
amplitude $A$ is shown in Fig. 3. The slosh excitation amplitude
$A$ is compared to the axial r.m.s. size $\sigma_{\sf rms}$ of the
$m_f=9/2$ component. The quantity $(A/\sigma_{\sf rms})^2$,
measured after expansion, is proportional to the ratio of the
slosh excitation energy to the internal kinetic energy per
particle. Data in Fig. 3 were taken with a gas at
$T/T_{F,9/2}=0.4$ and one at $T/T_{F,9/2}=1.2$, corresponding to
the quantum and classical regimes.  For the non-degenerate gas we
observe no non-linearity and the damping time is independent of
the excitation amplitude over a wide range of excitation energy.
In contrast, in the quantum regime we observe a strong amplitude
dependence to the damping time. In order to minimize the impact of
this non-linearity the data with $T/T_F<1$ in Fig. 2 were taken
with $(A/\sigma_{\sf rms})^2<0.01$.

\begin{figure}
\begin{center}\epsfxsize=3 truein \epsfbox{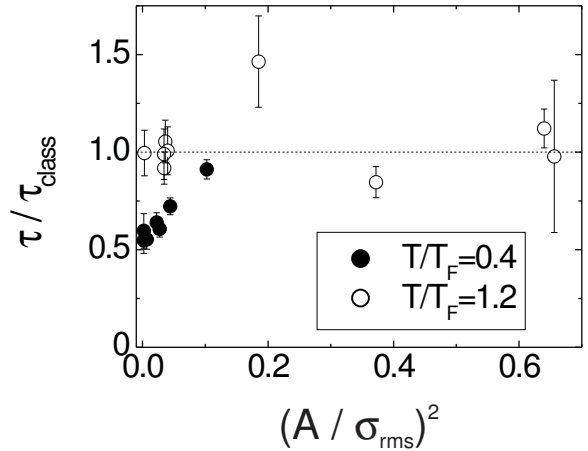}
\end{center}
\caption{Non-linear excitation dynamics.  Data for a
non-degenerate gas at $T/T_F=1.2$ are compared to data for a gas
in the quantum regime at $T/T_F=0.4$.  For the high temperature
data the measured damping time $\tau$ matches the classical
expectation (dashed line) $\tau_{\sf class}$ over a wide range of
excitation amplitude $A$. In the quantum regime, however, the
Pauli blocking effect on $\tau$ is strongly reduced as the
relative amplitude $A/\sigma_{\sf rms}$ increases. \label{fig3}}
\end{figure}

The observed non-linearity presumably comes from the energy
dependence of the Pauli blocking effect on collisions.  A large
excitation amplitude increases the average collision energy, and
therefore reduces the impact of Pauli blocking. For example, a
collision in which an atom gains an energy which is very small
compared to the Fermi energy, $E_F=k_bT_F$, will be strongly
suppressed in a quantum degenerate gas.  Conversely, a collision
in which an atom gains an energy that is larger than $E_F$ will
not be suppressed at all.  We find that the effect of Pauli
blocking on $\tau$ is significantly reduced for relatively small
slosh amplitude.  This non-linear behavior could provide a probe
of the Fermi sea structure through the energy dependence of Pauli
blocking.  Future work along these lines would require a
quantitative model for comparison with the observed effect.

In conclusion, we have used a  collective spin excitation in the
hydrodynamic regime to probe the interplay between interactions
and quantum statistics in a Fermi gas of atoms.  We measure an
increased damping rate compared to the classical expectation at
low temperature, and observe non-linear behavior in the quantum
regime.  If future experiments are able to cool a Fermi gas of
atoms to even lower $T/T_F$, studies similar to those presented
here could be used to explore the crossover from hydrodynamic to
collisionless behavior due to Pauli blocking in the degenerate
gas.  Excitation dynamics could also be used to investigate the
predicted phase transition to a Cooper-paired superfluid state in
a two-component Fermi gas of atoms
\cite{Baranov2000,Bruun2000b,Minguzzi2001}.

This work is supported by the National Science Foundation, the
Office of Naval Research, and the National Institute of Standards
and Technology.  The authors would like to express their
appreciation for useful discussions with C.E. Wieman and E.A.
Cornell and for work by S. B. Papp.


\begin{references}
\bibitem{note} Two of the lowest $\Gamma_{\sf coll}$ data points in
Fig. 1 were taken using a weaker trapping potential (135 Hz radial
frequency and 19.8 axial frequency for an $m_f=9/2$ atom).
\end{references}
\end{document}